\documentclass[aps,prl,twocolumn,floatfix,showpacs,citeautoscript,superscriptaddress,longbibliography,hyperlinks]{revtex4-2}

\usepackage[a4paper,left=15mm,top=22mm,right=15mm,bottom=22mm]{geometry}

\usepackage[colorlinks=true,citecolor=blue]{hyperref}
\usepackage[english]{babel}
\usepackage[utf8]{inputenc}
\usepackage{hyperref}
\usepackage{amsmath}
\usepackage{amssymb}
\usepackage{amsfonts}
\usepackage{graphicx}
\usepackage{siunitx}
\usepackage{color}

\newcommand{\revision}[1]{{{#1}}}

\begin{document}
\title{Real-Time Detection and Control of Correlated Charge Tunneling in a Quantum Dot}

\author{Johannes C. Bayer}
\affiliation{Institut für Festkörperphysik, Leibniz Universität Hannover, Hannover, Germany}
\affiliation{Physikalisch-Technische Bundesanstalt, 38116 Braunschweig, Germany}

\author{Fredrik Brange}
\affiliation{Department of Applied Physics, Aalto University, 00076 Aalto, Finland}

\author{Adrian Schmidt}
\affiliation{Institut für Festkörperphysik, Leibniz Universität Hannover, Hannover, Germany}
\affiliation{Institut für Technikfolgenabschätzung und Systemanalyse, Karlsruher Institut für Technologie, Karlsruhe, Germany}

\author{Timo Wagner}
\affiliation{Institut für Festkörperphysik, Leibniz Universität Hannover, Hannover, Germany}

\author{Eddy P. Rugeramigabo}
\affiliation{Institut für Festkörperphysik, Leibniz Universität Hannover, Hannover, Germany}

\author{Christian Flindt}
\affiliation{Department of Applied Physics, Aalto University, 00076 Aalto, Finland}
\affiliation{RIKEN Center for Quantum Computing, Wakoshi, Saitama 351-0198, Japan}

\author{Rolf J. Haug}
\affiliation{Institut für Festkörperphysik, Leibniz Universität Hannover, Hannover, Germany}

\begin{abstract}
\revision{We experimentally demonstrate the real-time detection and control of correlated charge tunneling in a dynamically driven quantum dot. Specifically, we measure the joint distribution of waiting times between tunneling charges and show that the waiting times for holes may be strongly correlated due to the periodic drive and the Coulomb interactions on the dot, although the electron waiting times are not. Our measurements are in excellent agreement with a theoretical model that allows us to develop a detailed understanding of the correlated tunneling events. We also demonstrate that the degree of correlations can be controlled by the  drive. Our experiment paves the way for systematic real-time investigations of correlated electron transport in low-dimensional nanostructures.}
\end{abstract}

\maketitle

\begin{figure}[b!]
	\centering
\includegraphics[width=0.9\columnwidth]{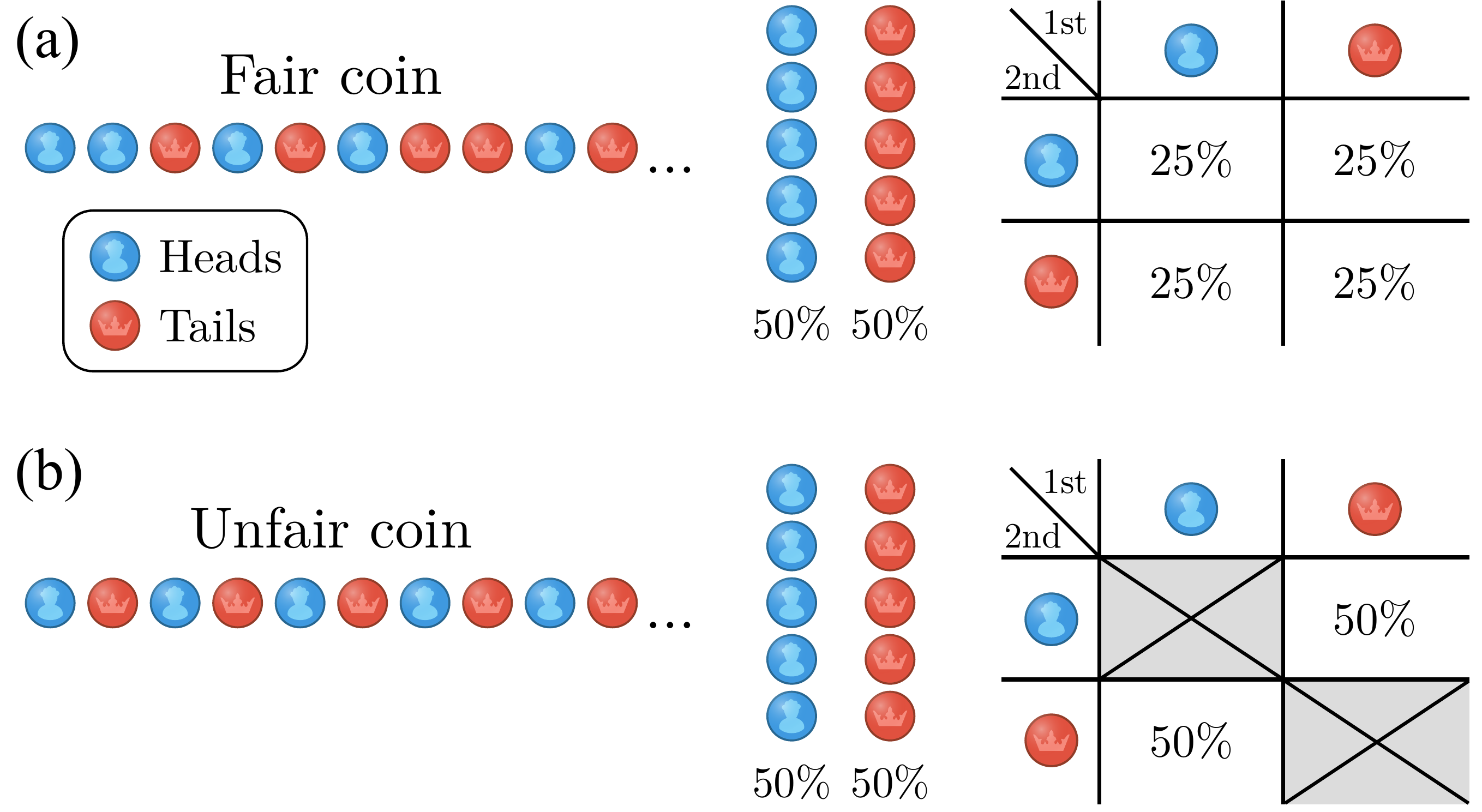}
\caption{Correlations in random processes. (a) Tossing a fair coin leads to a random sequence of heads and tails. One half of the tosses yield heads and the other half tails. Also, there are no correlations between the tosses, such that heads and tails are equally likely to be followed by heads or tails. (b)  One can instead consider an unfair coin that produces an alternating sequence of heads and tails. Also in that case, one half of the tosses yields heads and the other half tails. However, in this case, there are correlations between the coin tosses, since heads is always followed by tails and vice versa.}
	\label{fig:1}
\end{figure}

\emph{Introduction.---} Correlations in random processes are ubiquitous in nature and are important across a wide range of sciences and technologies~\cite{VanKampen2007}. A prominent example is that of entanglement in quantum physics, where the measurement on one particle can affect a subsequent measurement on another particle, even if they are far apart~\cite{Horodecki2008}. Correlations can also occur due to Coulomb interactions in nanostructures as the tunneling of one electron blocks the tunneling of the next one, which is visible in  noise measurements~\cite{Blanter2000}. The significance of correlations can best be appreciated with a simple example as in Fig.~\ref{fig:1}: If a fair coin is tossed, it has an equal chance of landing on either side up, and there are no correlations between the tosses with each coin toss being independent of the others. As a comparison, one can imagine an unfair coin that in each toss alternates between heads and tails. Such an unfair coin would yield heads half of the time and tails the other half, just as a fair coin. However, if one investigates the correlations between subsequent coin tosses, one would find that heads is always followed by tails, and tails is always followed by heads.

\begin{figure*}[t!]
	\centering
	\includegraphics[]{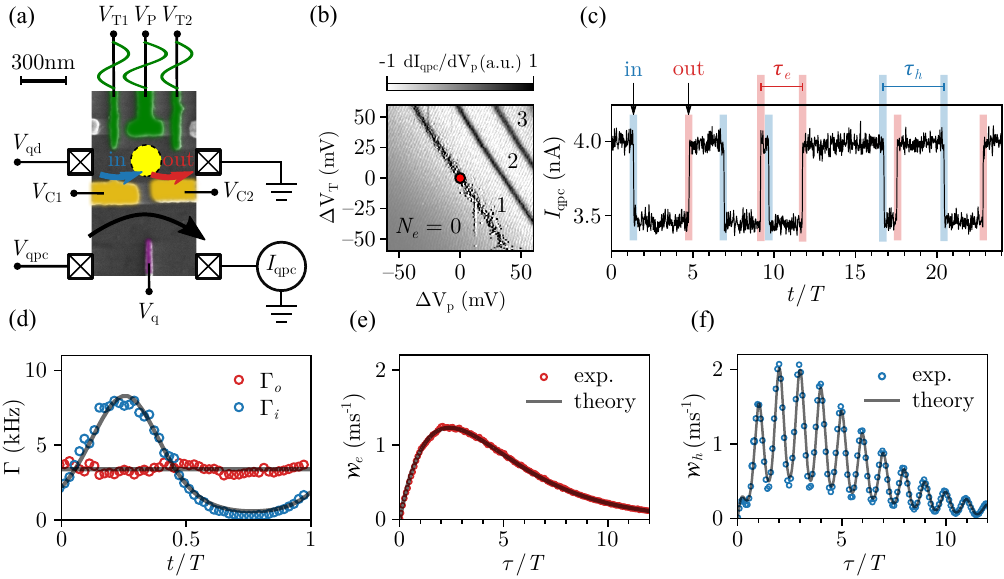}
\caption{Measurements of waiting times. (a) Colored SEM image of the device. A quantum dot (yellow circle) is formed by negative voltages on the gates \textbf{$V_\mathrm{T1,P,T2, C1, C2}$}. A quantum point contact (QPC) serves as a charge detector, which can be tuned by the gate $V_\mathrm{q}$. Ohmic contacts to the quantum dot are indicated by the crossed squares, and single electrons tunnel on and off the quantum dot from two electrodes. A small bias voltage $V_\mathrm{qd} = \SI{-1}{mV}$ drives a current through the quantum dot. \revision{The green gates $V_\mathrm{T1, P, T2}$} are used to modulate the tunneling rates periodically in time \revision{as described by Eq.~(\ref{eq:voltages})}. (b) Charge stability diagram of the device. The quantum dot is driven periodically \revision{with frequency $f=8$ kHz (and period $T=0.125$ ms)} around the operating point on the first charging line, where the charge states with $N_e = 0,1$ electrons on the dot are energetically degenerate  (red circle). Here, we use $\Delta V_{\mathrm{T}1}=\Delta V_{\mathrm{T}2}=\Delta V_{\mathrm{T}}$. (c) The time-dependent current through the QPC switches between two values corresponding to the number of electrons on the quantum dot, $N_e = 0,1$. The waiting times between tunneling events are indicated. (d) Time-dependent tunneling rates in and out of the quantum dot. \revision{The rate for tunneling out is roughly constant, while the rate for tunneling in depends on the periodic driving of the gate voltages via the exponential tunneling rate dependence shown in Eq.~(\ref{eq:tun_rates}).} (e)~Waiting time distribution for electrons tunneling out of the quantum dot. The experimental data are well captured by Eq.~(\ref{We}) corresponding to a device with constant tunneling rates. (f) Waiting time distribution for electrons tunneling into the quantum dot or, equivalently, holes tunneling out of the quantum dot. In this case, the experimental data are well captured by Eq.~(\ref{Wh}), which contains an oscillatory factor due to the time-dependent rate at which holes tunnel.}
	\label{fig:2}
\end{figure*}

In the context of quantum transport in nanostructures, several statistical tools have been used to describe the tunneling of charges~\cite{Nazarov2009}. In one approach, one considers the number of charges that have been transmitted through a mesoscopic conductor, which is known as the full counting statistics~\cite{Levitov1993,Levitov1996,Bagrets2003,Gustavsson2006,Flindt2009,Ubbelohde2012}. Alternatively, one may investigate the first-passage time distribution, which characterizes the time it takes for a given number of charges to have tunneled~\cite{Saito2016,Singh2019,Landi:2024}. In a complementary approach, one measures the distribution of waiting times between tunneling events~\cite{Brandes2008,Albert2011,Albert2012}. Photon waiting times have been investigated theoretically, however, they have rarely been measured since highly efficient detectors are needed~\cite{Vyas1988,Carmichael1989}. By contrast, measurements of electron waiting times have  become possible thanks to the development of accurate charge detectors. The time between tunneling charges can be described by a probability density, $\mathcal{W}(\tau)$, which has been measured in several experiments~\cite{Gorman2017,Jenei2019,Ranni2021,Brange2021,Hanze2021}. However, to identify correlations between waiting times, it is necessary to measure their joint distribution. The joint distribution is the probability density of first observing a waiting time of duration $\tau_1$ followed by a second one of duration $\tau_2$. Without correlations, it factorizes as~$\mathcal{W}(\tau_1,\tau_2)=\mathcal{W}(\tau_1)\mathcal{W}(\tau_2)$. Thus, to quantify correlations, one may consider the difference
\begin{equation}
    \Delta \mathcal{W}(\tau_1,\tau_2)=\mathcal{W}(\tau_1,\tau_2)-\mathcal{W}(\tau_1)\mathcal{W}(\tau_2),
\end{equation}
and the correlation ratio, which we define as
\begin{equation}
    \mathcal{R}(\tau_1,\tau_2)=\frac{\Delta\mathcal{W}(\tau_1,\tau_2)}{\mathcal{W}(\tau_1)\mathcal{W}(\tau_2)}.
    \label{def:R}
\end{equation}
This ratio is positive (negative) for positively (negatively) correlated waiting times and vanishes in the absence of correlations. Joint waiting time distributions have not been measured before, however, it has been predicted that \revision{waiting times may become correlated because of Coulomb interactions, the Pauli principle~\cite{Dasenbrook2015}, or superconductivity~\cite{Walldorf2018,Dutta2024}. In theory, it has also been found that a periodic drive can induce correlations~\cite{Hofer2016,aNote1}.}

\begin{figure*}[t!]
	\centering
	\includegraphics[]{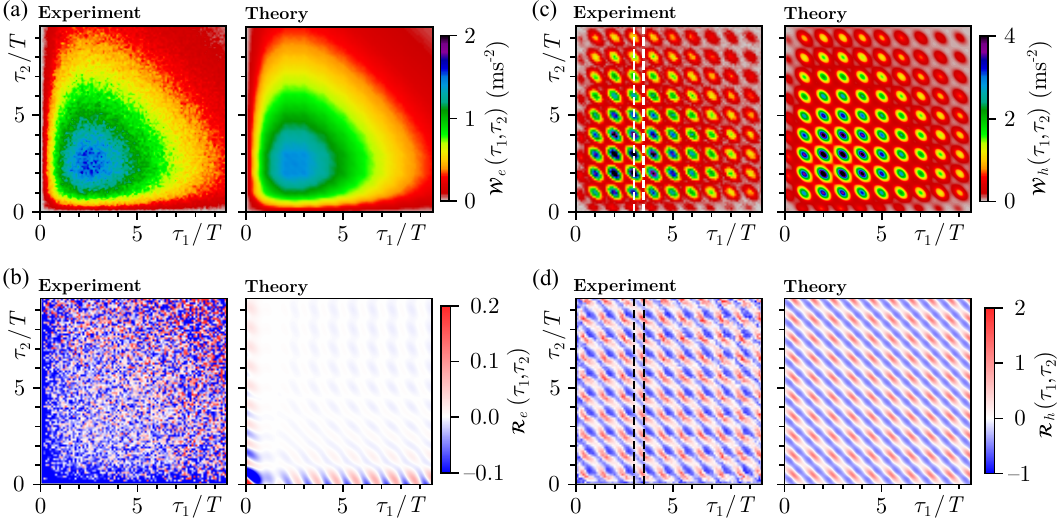}
 \caption{Correlated waiting times.  (a) Measurements of the joint waiting time distribution for the tunneling of electrons together with model calculations. (b) The corresponding correlation ratio shows that there essentially are no correlations between subsequent waiting times. (c) Measurements of the joint waiting time distribution for the tunneling of holes together with model calculations. (d) For the tunneling of holes, there are strong correlations between the waiting times.}
	\label{fig:3}
\end{figure*}

\revision{In this Letter, we report on the real-time detection and control of correlated charge tunneling in a dynamically driven quantum dot. To this end, we measure the joint distribution of electron and hole waiting times, and we find that the hole waiting times can be strongly correlated, although the electron waiting times are not. We develop a theoretical model that captures our measurements, and which allows us to show that the correlations between the hole waiting times arise because of the periodic drive and the Coulomb interactions between the tunneling charges. We also show that we can control the degree of correlations by changing the periodic drive, such that the electron waiting times also become correlated.}

\emph{Quantum dot device.---} Figure~\ref{fig:2}(a) shows our device, which is based on a GaAs/AlGaAs heterostructure with patterned metallic top gates and operated at a temperature of $1.5 \, \text{K}$. Negative voltages applied to the gate electrodes lead to the formation of a quantum dot in the two-dimensional electron gas about $110\, \text{nm}$ below the surface. To detect the electrons on the quantum dot, we use an adjacent quantum point contact (QPC), whose conductance changes as electrons tunnel in and out of the quantum dot \cite{Wagner2016,Bayer2017}. As shown in Fig.~\ref{fig:2}(b), we can map out the charge stability diagram of the dot by monitoring the conductance of the QPC as a function of the gate voltages. In the following, we operate the quantum dot close to the \revision{red} point, where the charge states with zero or one electrons on the dot are energetically degenerate. \revision{In addition, we apply periodic gate voltages of the form~\cite{Wagner2019}
	\begin{equation}
		\begin{bmatrix}
			\Delta V_\text{P}(t)\\
		 	\Delta V_\text{T}(t)
		\end{bmatrix}
		= 
		\begin{bmatrix}
			\Delta V^0_\text{P}\\ 
			\Delta V^0_\text{T}
		\end{bmatrix}
		\sin{\left(2 \pi f t \right)},
		\label{eq:voltages}
	\end{equation}
which defines the direction of the drive in the charge stability diagram, and $f =1/T$ is the driving frequency. We have performed measurements across a wide range of driving frequencies and directions. However, we focus on $f=8$~kHz, where an intricate interplay between the driving frequency and the tunneling rates is expected~\cite{Brange2021}. Moreover, for the amplitude and direction of the drive, we first take $\Delta V^0_\text{P}=3.2$ mV and 
$\Delta V^0_\text{T}=-9.5$ mV.}

As shown in Fig.~\ref{fig:2}(c), the time-resolved current in the QPC switches back and forth between two distinct values as electrons tunnel in and out of the dot. From such time traces, we can extract \revision{the tunneling rates of charges, which are affected by the periodic drive, because it modifies the tunnel barriers and the position of dot levels with respect to the Fermi levels of the electrodes. In Fig.~\ref{fig:2}(d), we show the extracted tunneling rates and see that the rate for tunneling out is roughly constant with $\Gamma_o\simeq 3.4$~kHz, because the change of the tunnel barrier for this particular drive is compensated by the shift of the dot levels.} On the other hand, the rate for tunneling in depends exponentially on the gate voltages as
\begin{equation}
    \Gamma_i(t)=\Gamma_ie^{\alpha \sin(2\pi f t)}/I_0[\alpha],
    \label{eq:tun_rates}
\end{equation}
where the parameter $\alpha\simeq 1.33$ characterizes the influence of the gate voltages on the tunneling rate~\cite{MacLean2007,Brange2021}. In the expression above, we have included the Bessel function, $I_0[\alpha]=\int_{0}^{2\pi}dx e^{\alpha\sin x}/2\pi$, so that $\Gamma_i \simeq 3.3 $~kHz is the average tunneling rate.  Those values are found by fitting Eq.~(\ref{eq:tun_rates}) to the experimental results in Fig.~\ref{fig:2}(d), and these rates form the basis for the modeling of our experiment.

\emph{Distribution of waiting times.---} From the time traces of the current in the QPC, we can extract the distribution of waiting times between subsequent tunneling events. In Fig.~\ref{fig:2}(e), we show the distribution of waiting times between electrons tunneling out of the quantum dot. This waiting time distribution depends only weakly on the periodic drive, and it is well-approximated by the expression
\begin{equation}
\mathcal{W}_e(\tau) \simeq \mathcal{W}_s(\tau)= \frac{\Gamma_i  \Gamma_o}{ \Gamma_i - \Gamma_o}(e^{-\Gamma_o \tau}-e^{- \Gamma_i \tau}).
\label{We}
\end{equation}
which is the distribution for a static device with constant tunneling rates~\cite{Brandes2008}. We obtain this expression by assuming that the tunneling rates are roughly constant between each tunneling event (see End Matter).

In Fig.~\ref{fig:2}(f), we show the distribution of waiting times between electrons tunneling into the quantum dot. We can equivalently think of these results as the waiting times between holes tunneling into the electrodes. For the hole waiting times, we find the expression\revision{
\begin{equation}
\mathcal{W}_h(\tau) \simeq \mathcal{W}_s(\tau) I_0[ \alpha g(\tau)]/I_0^2[\alpha]
\label{Wh}
\end{equation}
with $g(\tau)=2\cos(\pi f \tau)$}, which again is  given by the static distribution in Eq.~(\ref{We}), however, this time, it is multiplied by an oscillatory function, which accounts for the time-dependent rate at which holes tunnel out of the dot (see End Matter). This prediction agrees well with the measurements of the waiting time distribution in  Fig.~\ref{fig:2}(f).

\emph{Correlated waiting times.---} Although the results in Figs.~\ref{fig:2}(e) and \ref{fig:2}(f)  fully characterize the distributions of waiting times, they do not capture possible correlations between the waiting times. Similarly, in the example of coin tosses, both the fair and the unfair coin yield heads (or tails) half of the time. However, for the unfair coin, that does not capture the correlations between the coin tosses. To identify correlations between the waiting times, we need to consider their joint distribution and the correlation ratio in Eq.~(\ref{def:R}). Thus, in Fig.~\ref{fig:3}, we show measurements of the joint distributions of subsequent waiting times. Specifically, in Figs.~\ref{fig:3}(a) and \ref{fig:3}(b), we show the joint waiting time distribution for electrons and the correlation ratio. The experimental data are well captured by our model calculations, and the results resemble what one would expect for a static device with no correlations. Indeed, turning to the correlation ratio in Fig.~\ref{fig:3}(b), there are essentially no correlations, as the correlation ratio is vanishingly small and mostly dominated by experimental noise. Theoretically, we also expect that the correlations are small. If we assume that the tunneling rates are constant in between tunneling events, we predict a vanishing correlation ratio,
\begin{equation}    \mathcal{R}_e(\tau_1,\tau_2)\simeq 0,
    \label{eq:R1}
\end{equation}
which is in good agreement with the measurements.

\begin{figure}[t!]
	\centering
	\includegraphics[width=0.95\columnwidth]{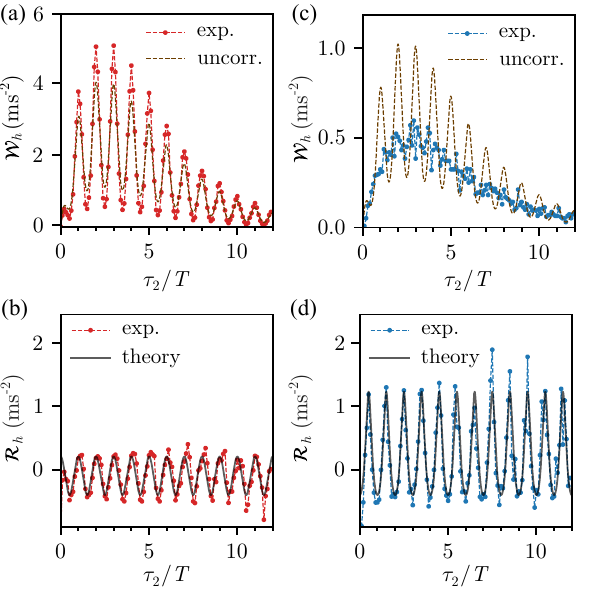}
	\caption{Joint distributions and correlation ratio. (a)~Joint distribution  along the first dashed line in Fig.~\ref{fig:2}(c) with $\tau_1=3T$. We also show the distribution without correlations, given by the single waiting time distribution.  (b) Correlation ratio  along the first dashed line in Fig.~\ref{fig:2}(d) with the theory given by Eq.~(\ref{eq:R2}). (c) Joint distribution along the second dashed line in Fig.~\ref{fig:2}(c) with $\tau_1=3.5T$. We also show the distribution without correlations, given by the single waiting time distribution. (d) Correlation ratio  along the second dashed line in Fig.~\ref{fig:2}(d) with the theory given by Eq.~(\ref{eq:R2}).} 
	\label{fig:4}
\end{figure}

In Fig.~\ref{fig:3}(c), a different picture emerges for the joint waiting time distribution of holes. The joint distribution exhibits an oscillatory behavior due to the time-dependent rate at which holes tunnel, and it is well-captured by our model calculations. Moreover, the correlation ratio in Fig.~\ref{fig:3}(d)  reveals that there are strong correlations between the waiting times, which can be both positive (red) and negative (blue), such that the measurement of a particular waiting time will likely be followed by a waiting time that is longer or shorter than one might have expected~\cite{Hofer2016,aNote1}. We can again calculate the correlation ratio with the simplification that the rates are constant in between tunneling events. We then find
\begin{equation}
	\revision{
\mathcal{R}_h(\tau_1,\tau_2)\!\simeq \! \frac{I_0[\alpha]I_0\big[\alpha\sqrt{3\!+\!g(2\tau_1)\!+\!g(\tau_1)g(\tau_1\!+\!2\tau_2)}\big]}{I_0[ \alpha g( \tau_1)]I_0[ \alpha g(\tau_2)]}\!-\!1,}
    \label{eq:R2}
\end{equation}
which \revision{captures the results in Fig.~\ref{fig:3}(d), and}, surprisingly perhaps, is independent of the average tunneling rates. 

In Fig.~\ref{fig:4}, we show the joint waiting time distribution and the correlation ratio along the two lines indicated in Figs.~\ref{fig:3}(c) and \ref{fig:3}(d). Along the first line, \revision{where the first waiting time is an integer number of periods, $\tau_1=3T$}, the joint distribution in Fig.~\ref{fig:4}(a) oscillates, and it resembles the distribution for uncorrelated waiting times. For that reason, the correlation ratio in Fig.~\ref{fig:4}(b) is small with periods of positive and negative correlations. By contrast, \revision{along the second line, where the first waiting time is a half-integer times the period, $\tau_1=3.5T$, the joint distribution in Fig.~\ref{fig:4}(c) shows that the measurement outcome for the first waiting time has a strong effect on the following waiting time, which loses its synchronization with the drive. Indeed, from Eqs.~(\ref{eq:holeWTD}) and (\ref{eq:holeWTD_halfInt}) of the End Matter, we see that the oscillatory dependence on the second waiting time gets washed out as the first waiting time changes from an integer number of periods to a half-integer number. This loss of synchronization leads to large deviations from the uncorrelated distribution and thereby to large oscillations in the correlation ratio as seen in Fig.~\ref{fig:4}(d).}

\begin{figure}[t!]
	\centering
	\includegraphics[width=0.95\columnwidth]{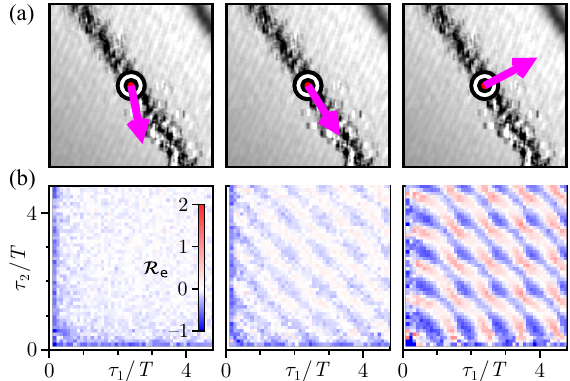}
	\caption{Control of correlations. (a) Charge stability diagram showing the direction of the periodic drive with respect to the single-electron charging line. (b) Corresponding correlation ratios for the electron waiting times, which increase as we change the direction of the periodic drive. The left panel corresponds to the results in Fig.~\ref{fig:3}(b), however, using a different scale and a different range of waiting times.}
	\label{fig:5}
\end{figure}

\emph{Control of correlations.---} Finally, we show that we can control the degree of correlations between the waiting times. To this end, we change the direction of the periodic drive as indicated in Fig.~\ref{fig:5}(a). The corresponding correlation ratios for electrons are shown in Fig.~\ref{fig:5}(b), and we see how the correlations increase as we change the drive and eventually apply the drive in a perpendicular direction to the charge-degeneracy line. \revision{In that case, both tunneling rates are time-dependent (not shown) unlike in Fig.~\ref{fig:2}(d), where one of them is approximately constant. In combination,} these results demonstrate how the degree of correlations can be controlled by the drive. 

\emph{Conclusions.---} We have measured the joint distribution of waiting times between tunneling charges and identified strong correlations, which arise because of Coulomb interactions and an external periodic drive. We have also shown how the degree of correlations can be controlled by changing the periodic drive. We have performed real-time measurements of the tunnel processes in a low-dimensional structure, however, future experiments may seek to detect correlations between charge pulses that propagate in elongated electronic wave guides~\cite{Bocquillon:2013}. Our experiment establishes a systematic approach for real-time investigations of correlated quantum transport, which is not limited to electrons and holes, but may also be extended to other discrete quanta such as single photons~\cite{Aharonovich:2016} or phonons~\cite{Cohen:2015}. \revision{Also, with the development of faster charge detectors, it is  conceivable that our approach can be implemented at higher driving frequencies, which would be important for metrological applications.}

\acknowledgments
\emph{Acknowledgments.---} This work was supported by the Deutsche Forschungsgemeinschaft (DFG, German Research Foundation) under Germany's Excellence Strategy -- EXC 2123
QuantumFrontiers -- 390837967, the State of Lower Saxony of Germany via the Hannover School for Nanotechnology, Jane and
Aatos Erkko Foundation, Research Council of Finland through the Finnish Centre of Excellence in Quantum
Technology (project number 352925), and Japan Society for the Promotion of Science through an Invitational Fellowship for Research in Japan.

\section*{End Matter}
\emph{Experiment.---} Our device is based on a two-dimensional electron gas (2DEG), which is formed in a GaAs/AlGaAs heterostructure approximately $\SI{110}{nm}$ below the surface. A quantum dot is formed in the 2DEG by applying negative voltages to metallic gates, which are patterned on top of the heterostructure by electron beam lithography.
The measurements were carried out in a $^4$He cryostat at $\SI{1.5}{K}$ in a dc transport setup. A $\SI{100}{kHz}$ bandwidth low-noise FEMTO transimpedance amplifier (\revision{$\SI{50}{MVA^{-1}}$} gain) was used to amplify the QPC current at room temperature. Signals were applied and recorded with a sampling rate of $\SI{400}{kHz}$ using an ADwin Pro2 real-time controller ($\SI{1}{GHz}$ ADSP T12).

The rates for tunneling in and out of the dot were extracted from the time-traces of the QPC current~\cite{Wagner2019}.
For small time bins $\Delta t$, the rates can be found as 
\begin{equation}
	\Gamma_\text{o}(t) = \frac{1}{\Delta t} \frac{N_\text{out}(t)}{N_1(t)},\quad
	\Gamma_\text{i}(t) = \frac{1}{\Delta t} \frac{N_\text{in}(t)}{N_0(t)},
\end{equation}
where $N_{\text{in/out}}(t)$ is the number of tunneling events in and out of the dot during the time $\Delta t$, and $N_{0/1}(t)$ measures how long the dot was occupied by zero or one electron.

\emph{Model.---} The system is described by the rate equation 
\begin{equation}
	\frac{d}{dt}|p(t)\rangle = \mathbf{L}(t)|p(t)\rangle,    
\end{equation}
where the vector $|p(t)\rangle = [p_0(t),p_1(t)]^T$ contains the probabilities for the quantum dot to be empty or occupied. The dynamics is governed by the rate matrix 
\begin{equation}
	\mathbf{L}(t)=
	\begin{bmatrix}
		-\Gamma_i(t) & \Gamma_o(t) \\
		\Gamma_i(t) & -\Gamma_o(t)
	\end{bmatrix},
\end{equation}
where $\Gamma_i(t)$ and $\Gamma_o(t)$ are the time-dependent tunneling rates. For the results in Figs.~\ref{fig:2}--\ref{fig:4}, the rate for tunneling in is given by Eq.~(\ref{eq:tun_rates}), while the rate for tunneling out to a good approximation is constant. We also introduce the jump operators $\mathbf{J}_i(t)$ and $\mathbf{J}_o(t)$, which describe the tunneling of electrons in and out of the quantum dot, 
\begin{equation}
	\mathbf{J}_i(t)=
	\begin{bmatrix}
		0 & 0 \\
		\Gamma_i(t) & 0
	\end{bmatrix},\quad
	\mathbf{J}_o(t)=
	\begin{bmatrix}
		0 & \Gamma_o(t) \\
		0 & 0
	\end{bmatrix}.
\end{equation}
The distribution of waiting times between electrons ($x=o$)  or holes ($x=i$) can then be expressed as
\begin{equation}
	\mathcal{W}_x(\tau) =\int_0^T dt \frac{\langle 1 |\mathbf{J}_x(t+\tau)\mathbf{U}_x(t+\tau,t)\mathbf{J}_x(t)|t\rangle}{\langle 1 |\mathbf{J}_x(t)|t\rangle} P_x(t),
	\label{eq:wtdtheo}
\end{equation}
where we have defined the time-evolution operator 
\begin{equation}
	\mathbf{U}_x(t_1,t_0) = \mathcal{T} \left\{e^{\int_{t_0}^{t_1} dt \left[\mathbf{L}(t)-\mathbf{J}_x(t)\right]}\right\},
\end{equation}
which excludes tunneling events of type $x=o,i$, and   $\mathcal{T}$ is the time-ordering operator. We have also defined the cyclic state at time $t$, given by the eigenproblem $\mathbf{U}_x(t+T,t)|t\rangle=|t\rangle$, together with $\langle 1| = (1,1)$. Moreover, the probability density for the time of a tunneling event reads 
\begin{equation}
	P_x(t) = \langle 1 |\mathbf{J}_x(t)|t\rangle/\int_0^T dt'\langle 1 |\mathbf{J}_x(t')|t'\rangle.
\end{equation}
Equation~(\ref{eq:wtdtheo}) can be understood by considering two tunneling events that are separated by the waiting time~$\tau$ with no other events in between. We then integrate over all possible times within the period of the drive that the first tunneling event could have occurred.

Following a similar line of arguments, the joint distribution of subsequent waiting times reads
\begin{widetext}
	\begin{equation}
		\mathcal{W}(\tau_1,\tau_2 ) =\int_0^T dt \frac{\langle 1 |\mathbf{J}_x(t+\tau_1+ \tau_2) \mathbf{U}_x(t+\tau_1+ \tau_2,t+\tau_1)\mathbf{J}_x(t+\tau_1)\mathbf{U}_x(t+\tau_1,t)\mathbf{J}_x(t)|t\rangle}{\langle 1 |\mathbf{J}_x(t)|t\rangle} P_x(t).
		\label{eq:jointwtdtheo}
	\end{equation}
\end{widetext}
This expression describes three tunneling events that are separated by the waiting times $\tau_1$ and $\tau_2$ with no other tunneling events occuring in between. Again, we integrate over all times that the first tunneling event could have occurred. In the main text, we evaluate Eqs.~(\ref{eq:wtdtheo}) and (\ref{eq:jointwtdtheo}) numerically
	to obtain the theoretical results in Figs.~\ref{fig:2} and~\ref{fig:3}. However, below we evaluate them analytically using a simple approximation that is valid at high driving frequencies.
	
	\emph{Fast driving.---} For high driving frequencies, the time-evolution operator $\mathbf{U}_x(t+\tau,t)$ can be approximated by
	\begin{equation}
		\mathbf{U}_x(t+\tau,t) \simeq e^{ \left( \mathbf{\bar L}- \mathbf{\bar J}_x \right)\tau},
	\end{equation}
	where $\mathbf{\bar L}$ and $\mathbf{\bar J}_x$ are the rate matrix and the jump operator with the time-dependent rates $\Gamma_i(t)$ and $\Gamma_o(t)$ replaced by their time-averages, $\Gamma_i = \int_0^T \Gamma_i(t)dt/T $ and $\Gamma_o = \int_0^T \Gamma_o(t)dt/T $.
	From Eq.~\eqref{eq:wtdtheo}, we then find 
	\begin{equation}
		\mathcal{W}_x(\tau) \simeq \Gamma_{\bar x}\frac{e^{- \Gamma_{\bar x} \tau}-e^{- \Gamma_x \tau} }{ \Gamma_x - \Gamma_{\bar x}} \frac{\int_0^T \Gamma_x(t+\tau)\Gamma_x(t) dt}{\int_0^T \Gamma_x(t) dt}.
	\end{equation}
	Plugging in the expressions for the time-dependent rate in Eq.~\eqref{eq:tun_rates}, we obtain Eqs.~\eqref{We} and \eqref{Wh} in the main text.
	
	Using the same approximation as above, we find the joint waiting time distributions as
	\begin{equation}
		\begin{split}
			\mathcal{W}_x(\tau_1,\tau_2) \simeq &\Gamma_{\bar x}^2\frac{e^{- \Gamma_{\bar x} \tau_1}-e^{- \Gamma_x \tau_1} }{ \Gamma_x - \Gamma_{\bar x}}\frac{e^{- \Gamma_{\bar x} \tau_2}-e^{- \Gamma_x \tau_2} }{ \Gamma_x - \Gamma_{\bar x}} \\
			&\times\frac{\int_0^T \Gamma_x(t+\tau_1+\tau_2)\Gamma_x(t+\tau_1)\Gamma_x(t) dt}{\int_0^T \Gamma_x(t) dt}.
		\end{split}
		\end{equation}
	With a constant rate for tunneling out, $\Gamma_o(t) = \Gamma_o$, and the time-dependent rate in Eq.~\eqref{eq:tun_rates}, we find
	\begin{equation}
		\mathcal{W}_e(\tau_1,\tau_2) \simeq \mathcal{W}_s(\tau_1)\mathcal{W}_s(\tau_2),
	\end{equation}
	and
	\begin{widetext}
	\begin{equation}
		\mathcal{W}_h(\tau_1,\tau_2) \simeq \mathcal{W}_s(\tau_1)\mathcal{W}_s(\tau_2)I_0\left(\alpha\sqrt{3+2\cos(2\pi f \tau_1)+4\cos(\pi f \tau_1)\cos(\pi f[\tau_1+2\tau_2])}\right)/I^3_0(\alpha),
		\label{eq:holeWTD}
	\end{equation}
	which then lead us to Eqs.~(\ref{eq:R1}) and (\ref{eq:R2})~for the corresponding correlation ratios. \revision{In general, the joint waiting time distribution for holes display an oscillatory behavior given by the driving frequency. However, for $\tau_1=(n+1/2)T$, we find}
		\begin{equation}
			\mathcal{W}_h((n+1/2)T,\tau_2) \simeq \mathcal{W}_s((n+1/2)T)\mathcal{W}_s(\tau_2)/I^2_0(\alpha),
			\label{eq:holeWTD_halfInt}
		\end{equation}
Thus, going from $\tau_1 = nT$ to $\tau_1 = (n+1/2)T$, the oscillatory dependence on $\tau_2$ is suppressed as seen in Fig.~\ref{fig:4}(a)~and~\ref{fig:4}(c).
\end{widetext}

\begin{thebibliography}{32}
\bibitem [{aNo()}]{aNote1}%
	\BibitemOpen
	\href@noop {} {\bibinfo {title} {\revision{As an example, one may consider busses that are scheduled to depart periodically. If one bus leaves on time, but the second one is delayed, the waiting time between them will be longer than the period. However, the subsequent waiting time between the second and the third bus will then likely be shorter than the period. Thus, a long waiting time is likely followed by a short one and vice versa.}}}
			\BibitemShut {Stop}%
\end{thebibliography}
\end{document}